# The Photothermal Effect in Interferometers

Shanti Rao

*LIGO Laboratory, Division of Physics, Math, and Astronomy*

*California Institute of Technology, Pasadena, CA 91125, USA.*

Abstract

We have measured the photothermal effect in a single cross-polarized interferometer at audio frequencies (5 Hz - 4 kHz). In a Fabry-Perot interferometer, light in one polarization is chopped to periodically heat the interferometer mirrors, while light in the orthogonal polarization measures the mirror length changes. Tests of a polished solid metal mirror show good agreement with relevant proposed theories by Braginsky *et al.* [Braginsky 1999] and Cerdonio *et al.* [Cerdonio 2001] describing uncoated optics.

## 1 Introduction

Optical interferometers (IFOs) can be used to measure small distance changes and are used for precision measurements such as in gravitational wave (GW) detectors. We are concerned with noise in these measurements driven by the light itself. Since available mirrors invariably absorb a small amount of the light that strikes them, photon shot noise and other intensity variations in light beams cause measurement errors, as photons are randomly converted to heat at the mirror surface. Assuming the only intensity fluctuations in the laser are from shot noise, photon-driven noise is believed to be an order of magnitude smaller than the standard quantum limit of sensitivity for LIGO and similar detectors. If, however, the lasers are *not* shot-noise limited, then these noise sources will limit the detectors' sensitivities.

Two expected problems in interferometric GW detectors are the photothermal effect [Braginsky 1999], in which the IFO mirrors expand and contract as heat is deposited into a mirror, and the photorefractive (dn/dT) effect [Braginsky 2000], in which optical heating changes the indices of refraction of multilayer dielectric mirrors, causing variations in the phase shift of the light reflected from a mirror.

The test masses in interferometric GW detectors consist of thick, transparent substrates with thin, multilayer dielectric mirrors on their surfaces. Pairs of test masses are arranged to make long Fabry-Perot interferometers, so that a strain in spacetime from a gravitational wave will move one end of the IFO relative to the other. If a laser beam is resonant with the IFO, the magnitude and phase of the beam reflected from the IFO will depend strongly on the motions of the mirror. A difficulty arises in that the gravitational waves act on the center of mass of a test mass, whereas the light senses only the position of the mirror on the mass's surface. Thus, surface effects like thermal expansion, dn/dT, and Brownian motion tend to set sensitivity limits for these instruments.

## 1.1 Photothermal noise

Photothermal noise is driven by laser intensity fluctuations such as shot noise. Since the manufacturer of the mirrors used in LIGO (Laser Interferometer Gravitational Wave Observatory) estimates that they absorb at least 1 ppm of the light that strikes them [1], intensity fluctuations in the laser cause length changes in the interferometer, as absorption converts light energy to heat energy, which drives thermal expansion of the mirror and substrate.

There are several theoretical models (see section 1.2) that describe photothermal noise in IFOs. Which of these are valid in a particular regime depends on the thermal expansion coefficient $\alpha$, the laser spot size $r_0$ (the radius at which the power is 1/e of the central value), and the thermal diffusion length $L_{therm} = \sqrt{\kappa / \rho C_P \omega}$. Here, $\omega$ is the frequency of the measurement, $\rho$ is the substrate density, $C_P$ is its heat capacity, and $\kappa$ is its thermal conductivity.

At high frequencies, when $L_{therm} \ll r_0$, an absorbed photon will create a temporary blister on the mirror. The $TEM_{00}$ laser beam, reflecting off the mirror, picks up an average phase shift which depends on the size, location, and age of the blister. Models of the photothermal effect assume a stochastic background of these blisters, distributed on the mirror surface according to the local intensity of the laser beam. In this regime, the magnitude of photothermal noise predicted by all the models is set by the factor, $\alpha / \rho C_P r_0^2 \omega$, neglecting constants and factors of order unity. This is the case for all the interferometric GW detectors presently active: LIGO, GEO, TAMA, and VIRGO all have spot sizes of several cm, while $L_{therm}$ for fused silica at 10 Hz is approximately 0.1 mm.

Table 1 lists common materials and their relative photothermal effect (neglecting Poisson's ratio). Given a fixed frequency and spot size, the final column contains the photothermal figure of merit for various materials. Fused silica, the substrate of choice for first-generation interferometric GW detectors, has the lowest photothermal response of common GW test mass materials, while the response for crystals and metals is quite high.

**Table 1** Thermal properties of potential mirror substrate materials. Values given are representative, and may depend on the alloy or direction with respect to a crystal axis.

| Material | $\alpha\ 10^{-6}/K$ | $\rho\ 10^3\,kg/m^3$ | $C_P$ J/kg K | $\alpha/\rho C \cdot 10^{-12}\,m^3/J$ |
|---|---|---|---|---|
| Al[a] (pure) | 23.1 | 2.7 | 897 | 9.5 |
| Al (6061)[b] | 23.1 | 2.7 | 963 | 8.9 |
| Cu[a] | 13.5 | 8.96 | 385 | 3.9 |
| Ti[a] | 8.6 | 4.51 | 523 | 3.6 |
| Be[a] | 11.3 | 1.85 | 1825 | 3.3 |
| GaAs[a] | 5.4 | 5.3 | 330 | 3.1 |
| Si[a] | 4.68 | 2.32 | 702 | 2.9 |
| $Al_3O_2$[c] Synthetic sapphire | 8.5 | 3.98 | 763 | 2.8 |
| $SiO_2$[d] Synthetic fused silica | 0.55 | 2.2 | 670 | 0.37 |

a. CRC Handbook, 77th ed.

b. http://www.matweb.com/SpecificMaterial.asp?bassnum=MA6016&group=General

c. http://www.crystalsystems.com/proptable.html

d. http://www.quartz.com/gedata.html

Previous investigations of the photothermal effect typically operate in the regime, $L_{therm} \approx r_0$, where the photothermal response of the material is most easily discernible [Bennis 1998]. Since models of the photothermal effect generally consider only heat deposited at the surface of a uniform, solid substrate, we verified the instrument and the theories with a solid Aluminum mirror.

Outside of the GW community (and particularly in the semiconductor industry), the photothermal effect is used for measuring the thermal conductivity of materials [Olmstead 1983], so research has focused on studying optically absorptive materials. There is much less work on transparent materials like the glasses and crystals used for making mirror substrates. An interferometer is a natural methodology to study the effect in mirrors, where the photothermal distance changes may be less than a nanometer.

De Rosa *et al.* [De Rosa 2002] have reported an observation consistent with the photothermal effect in low-absorption (0.5ppm) dielectric mirrors. Their study was done at low frequencies (10 mHz - 200 Hz), at which $L_{therm}$ is larger than the thickness of the mirror coating. They used two identical Fabry-Perot interferometers, whose mirrors had fused silica substrates.

## 1.2 Models

The Braginsky *et al.* [Braginsky 1999] model for the photothermal effect in interferometers assumes a large laser spot and a semi-infinite substrate. Subsequent work by Liu & Thorne [Liu 2000] identified corrections on the order of 5% that extend the Braginsky *et al.* model to finite-sized substrates with dimensions comparable to $r_0$. A further calculation was made Cerdonio *et al.* [Cerdonio 2001], who found an analytic solution covering low frequencies and small laser spots for the case of a semi-infinite substrate. As mirror substrate dimensions are typically much larger than thermal diffusion lengths, there is no need to correct this model in the manner of Liu & Thorne.

The Braginsky *et al.* photothermal model sets an upper limit on the photothermal effect in an interferometer mirror. It describes the asymptotic high-frequency response of a semi-infinite mirror, and assumes a negligibly small skin depth of the mirror. Averaged over a beam with a Gaussian intensity profile at normal incidence, the response function (in units of length) is:

$$F_B(\omega) = \sqrt{2}\alpha(1 + \sigma)\frac{P_{abs}(\omega)}{(\rho C_P \pi r_0^2)\omega} \tag{EQ 1}$$

where EQ 1 depends on the frequency $\omega$, the amount of power absorbed by the mirror $P_{abs}$, and the material properties of the mirror (Table 2):

**Table 2**

| | |
|---|---|
| $\alpha$ | coefficient of linear thermal expansion |
| $\sigma$ | poisson's ratio |
| $\rho$ | density |

**Table 2**

| | |
|---|---|
| $C_P$ | specific heat |
| $r_0$ | laser spot size (1/e of power) |
| $\kappa$ | thermal conductivity |

The Cerdonio *et al.* model [Cerdonio 2001] assumes the mirror is much larger than the spot size, and is valid at all frequencies:

$$F_C(\omega) = \sqrt{2}\alpha(1 + \sigma)\frac{P_{abs}(\omega)}{\pi\kappa}K(\omega\rho C r_0^2/\kappa)$$

(EQ 2)

where $\Omega \equiv \omega\rho C r_0^2/\kappa$ and

$$K(\Omega) = \int_0^\infty du \int_{-\infty}^\infty dv \left[\frac{u^2 e^{-u^2/2}}{(u^2 + v^2)(u^2 + v^2 + i\Omega)}\right]$$

(EQ 3)

In the limit of high frequencies (or short diffusion lengths), Cerdonio *et al.* have shown that $K(\Omega)$ approaches $1/\Omega$, and EQ 2 approaches EQ 1.

Both of these theories assume that the photothermal displacement is both caused and measured by collinear $TEM_{00}$ Gaussian laser beams. Heat deposited in a mirror will be concentrated at the center of the beam, where the light is most intense. Likewise, observations of length changes in the interferometer are dominated by the mirror surface displacement at the center of the beam.

## 2 Materials and Methods

### 2.1 Overview

Our goal is to measure the photothermal response of a single 'test' mirror in a Fabry-Perot interferometer. We split a 500 mW near-IR (1064nm) laser beam into two cross-polarized beams, *Force* and *Measure*, and then recombine and resonate them in a Fabry-Perot IFO. The *Force* beam is used to modulate the light power that strikes the test mirror, while the *Measure* beam detects length changes in the interferometer.

The *Force* beam is periodically interrupted with a chopping wheel, converting it to a pulsed beam with a 50% duty cycle. This enables a direct, lock-in measurement of the photothermal effect by recording IFO length changes at the frequency of the chopper. To measure length changes in the IFO, we use the Pound-Drever-Hall (PDH) [Drever 1983, Black 2001] optical heterodyne technique. This creates an 'error signal' $\varepsilon$, which, for a small range of frequencies, is a voltage proportional to the difference between the laser frequency and the cavity resonance frequency. The error signal measures the average displacement of the test mirror surface, weighted by the Gaussian beam profile.

This interferometer is made from two mirrors held apart by an aluminum spacer block. Its input mirror (24, Figure 1 on page 7) is a concave fused silica substrate with a dielectric mirror coating, a reflectivity of 95%, and a radius of curvature of 1 m. The output, or 'test', mirror (25) is a solid block of aluminum metal, diamond-turned and polished to obtain a flat, 89% reflective surface that absorbs approximately 1-2% of the light that strikes it. The reflectivity of the test mirror is somewhat uncertain, as oxidation and scratches can vary along its surface. The mode matching to the IFO is arranged so that the laser spot size at the input mirror is a few mm, while the spot size at the aluminum mirror is much smaller at 0.2mm. Since the photothermal effect strongly depends on the spot size, IFO length changes due to the photothermal effect will be dominated by the test mirror. The mode matching is largely insensitive to changes of a few cm in the lengths of the separate paths. The *Measure* beam, passing through more glass than the *Force* beam, has a 4 cm longer optical path length, but this does not significantly affect the coupling of the beams to the cavity.

The *Measure* beam travels through an electro-optical modulator (EOM), which phase shifts the beam to add radio frequency (RF) sidebands. When the laser is resonant with the IFO, the sidebands will be partially reflected and measured by an RF photodiode (RFPD). The high frequency output of the RFPD is demodulated with the PDH method to generate the error signal that measures changes in the cavity length. The error signal is also used with a feedback servo to keep the laser resonant with the IFO.

## 2.2 Materials

The equipment used is sketched in Fig. 1:

**Figure 1** Experiment layout. Shaded/colored lines represent laser beams, and solid black lines represent wires.

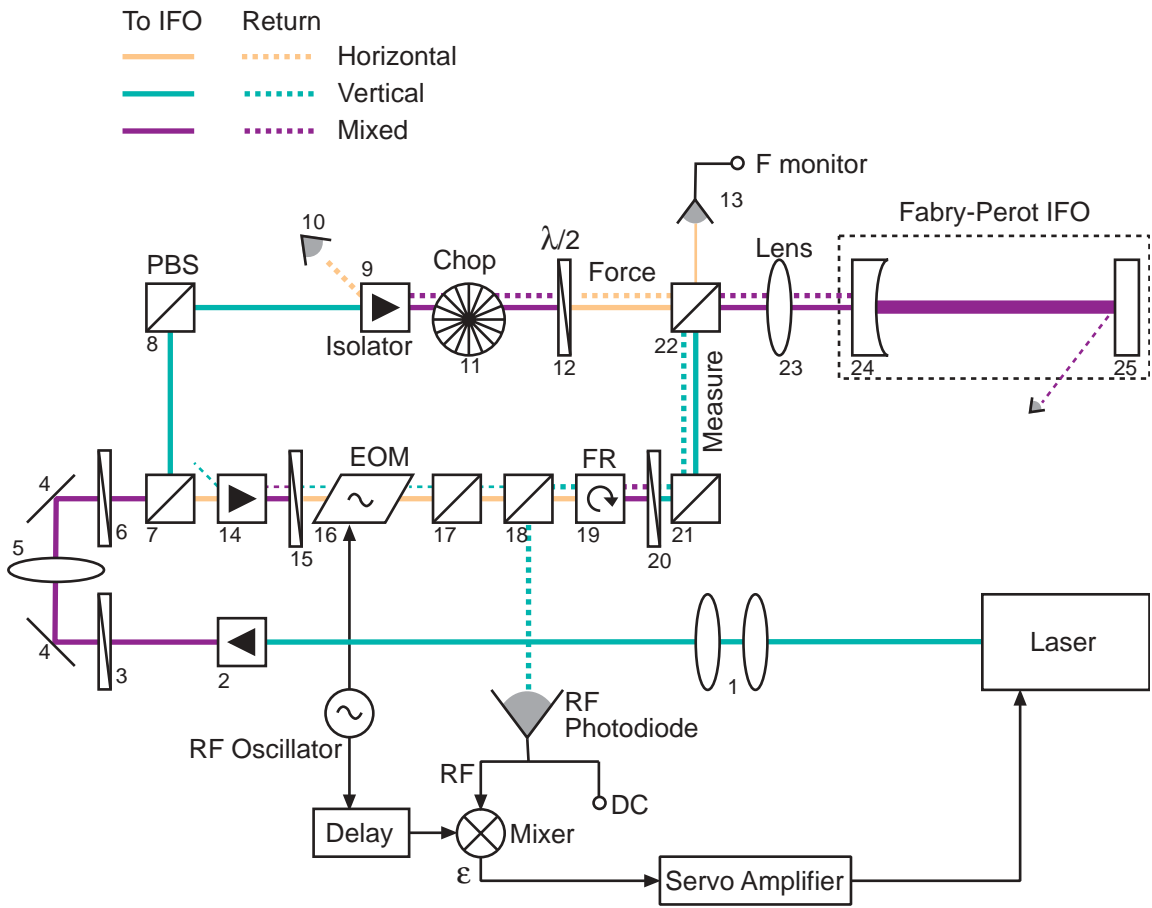

The vertically polarized beam leaves the laser, then travels through a mode-matching telescope (1) and a Faraday Isolator (2). The mode-matching telescope lenses are misaligned from the beam by approximately 3°, so that light reflected from them does not reenter the laser. The Faraday Isolator prevents light reflected off downstream optics from returning to the laser, and, as a side effect, rotates the beam polarization vector by 45°. A half-wave plate (3) restores vertical polarization prior to reflection off of steering mirrors (4). A mode-matching lens (5) images a beam waist at the modulator (16) in the *Measure* path and chopper (11) in the *Force* path.

After the steering mirrors, a half-wave plate (6) rotates the polarization of the beam to select the relative amounts of power that the first polarizing beamsplitter (PBS) (7) will direct into the *Force* and *Measure* beams. A PBS projects an incident beam's polarization onto its basis vectors: the horizontal component of the incident beam passes straight through, while the vertical component is reflected to the side. The rated extinction ratio of a PBS is 1000:1, so that we expect $\frac{1}{1000}$ of the horizontal and vertical components of the incident beam to exit the PBS along the wrong path.

In the *Force* path, a second PBS (8) turns the beam by 90º. Another Faraday Isolator (9) diverts back-reflections, which are monitored with a photodiode (10). The *Force* beam is chopped with a multi-slit, windmill-style chopper wheel (11), and another half-wave plate (12) rotates the polarization to horizontal. Then the *Force* beam is transmitted through the recombining beamsplitter (22), through the last mode-matching lens (23), and into the IFO. The horizontal component of the light that is reflected from the IFO will travel back along this path until it is removed by the Faraday Isolator (9).

In the *Measure* path, the horizontally polarized beam passes through another Faraday Isolator (14) and is rotated to again have horizontal polarization. A resonant electro-optic modulator (EOM) driven by a function generator (RF Oscillator) shifts 2.7% of the *Measure* energy into optical sidebands at $\pm 14.75\,\mathrm{MHz}$. The PBS (18) combines with the Faraday Rotator (19), half-wave plate (20), and PBS (21) to make a Faraday Isolator that will send light reflected from the IFO to the RF photodiode. The PBS (21) turns the *Measure* beam toward the final PBS (22), which recombines the square-wave *Force* beam and frequency-modulated *Measure* beam along the same path to resonate in the IFO. A small fraction of the *Force* beam is reflected by the final PBS. The *F monitor* photodiode detects this to track the position of the chopper wheel.

The high-frequency variation of the light intensity measured by the RF photodiode is demodulated with a mixer, driven by the same local oscillator as the EOM which created the sidebands. A delay line between the oscillator and mixer is used to maximize the response of the PDH error signal to length changes of the cavity. The error signal is amplified, recorded, filtered, and fed back to control the laser frequency.

The $\mathrm{TEM}_{00}$ resonant axis of the IFO is determined by the line normal to the surface of the test mirror that passes through the center of curvature of the input mirror. We adjust the tip and tilt of the optics to maximize the amount of light stored in the cavity on resonance, as we ensure that both *Force* and *Measure* beams are collinear after recombination, so that the *Force* beam heats the same spot that the *Measure* beam senses.

The Fabry-Perot interferometer itself is made of a fused silica input mirror (24) and a solid aluminum test mirror (25), both held in place by an aluminum spacer block.

The input mirror is a standard interferometer mirror from CVI Laser (part PR1-1064-10-1025-1.00CC) [3] with a commercial multilayer mirror coating formed by electron-beam deposition. Its flat surface has a commercial antireflection coating ($R < 0.25\%$). To reduce vibration, its concave, reflective surface abuts the end of the spacer block, and is pressed in place from the back under moderate pressure by an O-ring along its perimeter.

The test mirror is an aluminum cylinder (6061-T6 alloy) diamond-turned by Janos Technology Inc. (part A8010-101) and polished by CVI Laser. Of the light that it doesn't reflect, a good deal is scattered. It is difficult to precisely measure the ratio of absorbed to scattered light, as most of the scattered light reflects along the incident beam, but we expect that it absorbs 1-2% of 1 micron wavelength light at normal incidence. The nonreflective surface of this mirror is tapped with three holes and bolted to an aluminum plate which is itself bolted to the spacer block.

The IFO spacer block is a 12.5"x3"x3" rectangular brick of 6061 aluminum with holes drilled in opposite ends and a channel milled from the middle to allow light to travel through it. It is bolted to a pedestal which is clamped to the workbench. The inside of the channel is lined with layers of thermally insulating plastic and aluminum foil to prevent scattered light from being absorbed by the walls of the block. The channel is covered with a transparent acrylic lid to allow scattered light to escape, while preventing air currents from shaking the mirror surfaces.

A major challenge in designing this apparatus is controlling the cross-talk between the beams. Consider the *Force* beam, which is chopped, then impinges upon and is partially reflected from the IFO. If the *Force* beam heading into the IFO has a vertical component to its polarization vector, the process of separating and recombining the paths will create a Mach-Zender interferometer. Any component of *Force* with vertical polarization that reflects from the IFO will travel back along the *Measure* path and into the RF photodiode. This can cause a small systematic error in the measurement. We reduce this effect by carefully adjusting waveplates and by doubling beamsplitters, such as parts 17 and 21, which combine with their neighbors to allow an extinction ratio of $10^4$:1 to $10^5$:1. A detailed analysis of the effects of cross-talk is discussed in section 4.1.

## 2.3 Calibration

A summary of the measured properties of the interferometer is given in table 3. A description of the calibration procedures follows.

**Table 3**  Physical properties of the interferometer

| | | |
|---|---|---|
| Length | L | 12.5in = 0.318m |
| Wavelength | $\lambda$ | 1.064$\mu$ |
| Free spectral range | FSR = c / 2L | 447 MHz |
| PZT coefficient | $C_{PZT}$ | 6.3 MHz / $V_{PZT}$ |
| PZT bandwidth | $f_{PZT}$ | 55 kHz |
| resonance FWHM | $\Delta\nu_{FWHM}$ | 1.92 $V_{PZT}$ = 12.1 MHz |
| Finesse | FSR$/(\Delta\nu_{FWHM})$ | 37 |
| Visibility | | 50% |
| Input mirror (SiO$_2$) | 50 cm CC/PL | Reflectivity: 95% |
| Test mirror (Al) | PL | Reflectivity: ~89% (uncoated) |
| Demodulation phase | $\theta$ | 0.779 $\pi$ |
| Test spot size | $r_0$ | .20mm |
| PDH discriminant | $C_{PDH}$ | 0.33 |
| *Force* power | $P_F$ | 122 mW |
| *Measure* power | $P_M$ | 256 mW |

We use a 500 mW LightWave Model 126 laser [2], which uses a Nd:YAG Non-Planar Ring Oscillator to produce a continuous-wave beam with a wavelength of 1.064 microns. The laser frequency (and wavelength) may be changed slowly by a few MHz by varying the laser crystal temperature, or quickly by applying stress to the laser crystal with a PZT piezo-electric actuator bonded directly to the laser crystal. The fast (55 kHz) response of the PZT is used to lock the laser to the IFO (See page 12). The PZT tuning coefficient was measured with a high-finesse cavity.

The finesse of the cavity is determined by scanning it with a low-power laser beam (Fig. 2). An oscilloscope records the PDH signal (yellow), the reflected power measured at the RF photodiode (blue), light power scattered from the end mirror (a measure of the light stored in the cavity) (purple), and the control voltage sent to the PZT crystal that controls the laser frequency (green). The full-width at half-max of the scattered light peak is 12.1 MHz, implying a finesse of 37. The height of the dip in the reflected power trace indicates that the cavity's visibility is 50%.

**Figure 2** Yellow: Error signal, Blue: Reflected power of the *Measure* beam, Purple: Light scattered from IFO, Green: Laser frequency control (6.3 MHz/V). The reflected power and scattered light signals are measured by different types of photodiodes. The zero of the Reflected Power signal is at the bottom of the plot.

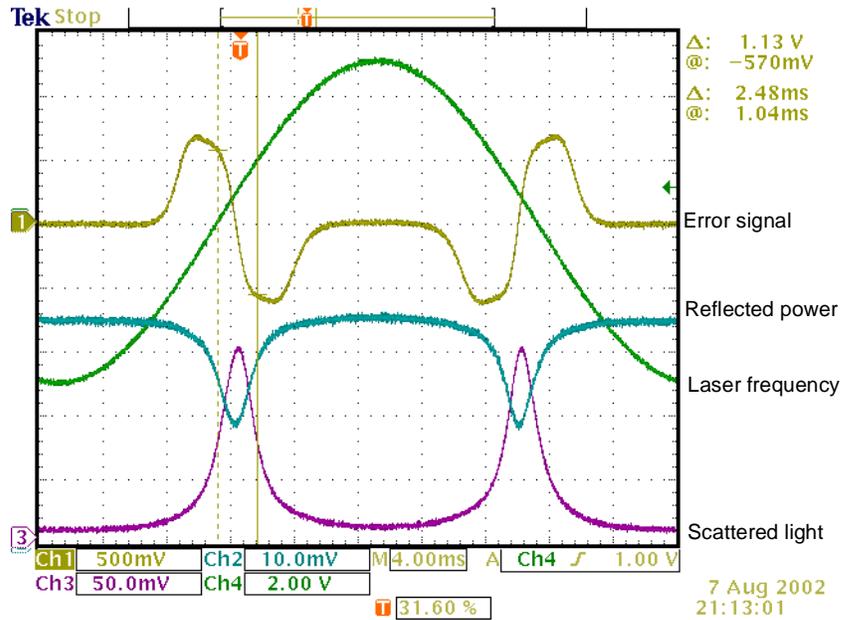

The same measurement can be used to verify the reflectivities of the interferometer mirrors, which are directly measured with a power meter. The PDH discriminant formula is compared to the measured PDH signal (Fig. 3), where the fit parameters are the vertical scale and demodulation phase ($\theta$, Table 3). The phase $\theta$ characterizes the delay box attached to the RF oscillator. This plot can also be used to measure the discriminant, but a more accurate method is described in section 2.4. The equation describing the PDH signal as a function of frequency is defined in [Black 2001] and in the Appendix (EQ 21).

**Figure 3** Pound-Drever-Hall (PDH) error signal. The input reflectivity is 95%, and the test mirror reflectivity is ~89%

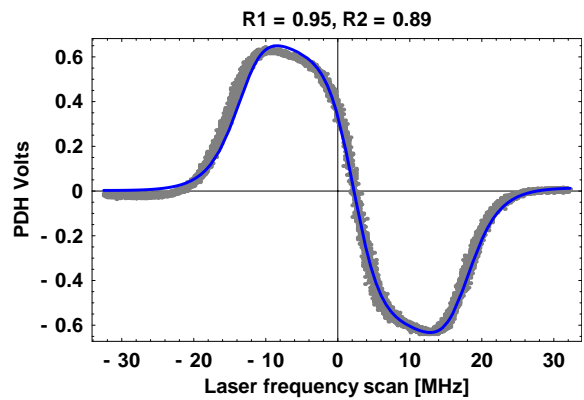

The spot size is calculated from the interferometer geometry, after the method described in

[Kogelnik 1966].

## 2.4 Operation

For the laser to reliably resonate in the IFO, a control servo (Fig. 4) continually measures the error signal and uses that information to adjust the laser frequency.

**Figure 4** Control servo topology to lock the laser frequency to the cavity

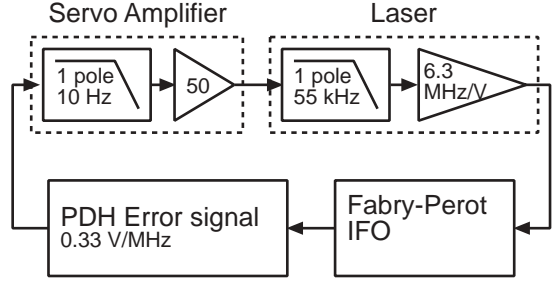

The servo is characterized by the dimensionless Open-Loop Transfer Function (OLTF), which includes the poles of the servo amplifier and the laser PZT:

$$\text{OLTF(f)} \ = \ \frac{\text{DC gain}}{(1 + if/(10 \text{ Hz}))(1 + if/f_{\text{PZT}})}$$

<div align="right">

**(EQ 4)**
</div>

The OLTF was measured directly with a spectrum analyzer. A two-parameter fit (Fig. 5) to the data was used to determine the PZT bandwidth ($f_{\text{PZT}}$, 55kHz) and the DC gain (104) of the OLTF. The servo amplifier gain (50) and pole (10 Hz) were measured directly. The PDH discriminant is the DC gain of the OLTF divided by the DC gains of the other components, $C_{\text{PDH}} \ = \ 104/(50 \cdot C_{\text{PZT}})$. This function agrees well with the magnitude response of the servo, but differs in phase by a few degrees from 400 Hz to 4 kHz. Multiplying the OLTF function with a shallow resonance $H_{\text{res}}$ more accurately predicts the phase; the function $\text{OLTF} \cdot H_{\text{res}}$ is obscured by the data points in the phase plot of Fig. 6. The zero (1050 Hz) and pole (1130 Hz) of the resonance function are obtained from a fit to the phase data:

$$H_{\text{res}} \ = \ \frac{1 + if/(1050 \text{ Hz})}{1 + if/(1130 \text{ Hz})}$$

<div align="right">

**(EQ 5)**
</div>

An effect of the control system is to suppress the measurements by a factor of the return difference function, $H_{R-D}$, with which the data are compensated in the analysis.

$$H_{R-D} = \frac{1}{1 + OLTF \cdot H_{res}}$$

(EQ 6)

**Figure 5** Magnitude response of the control servo

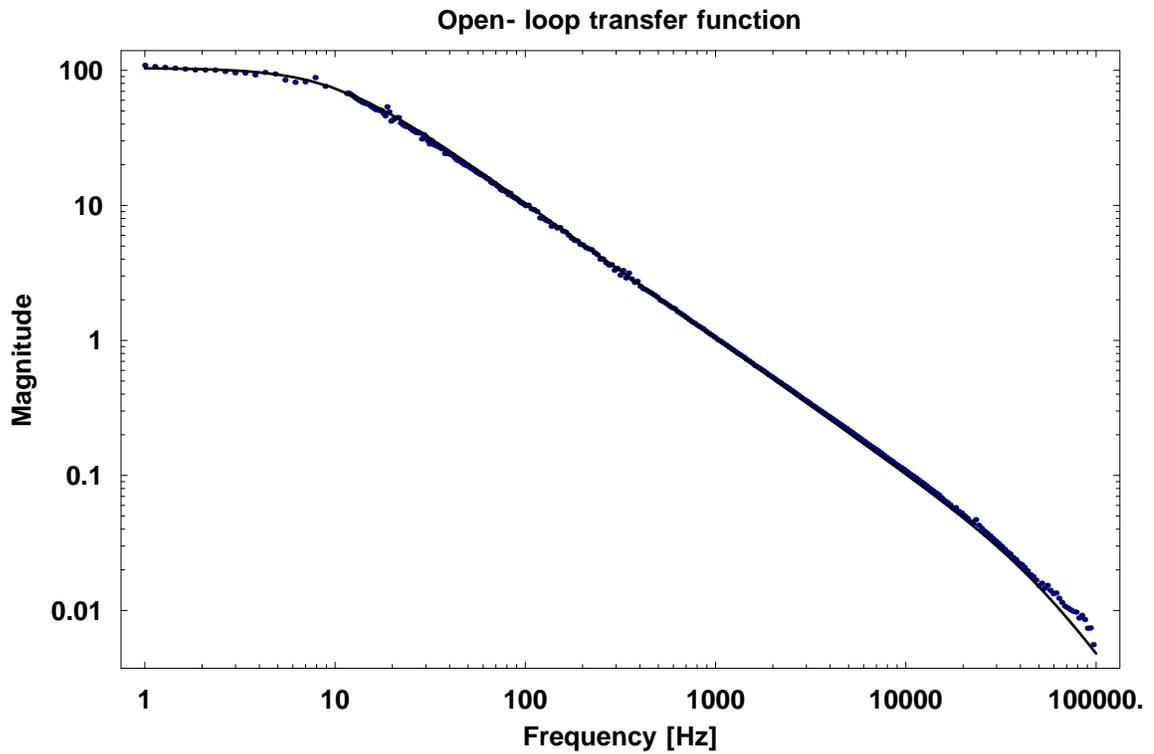

**Figure 6** Phase response of the control servo

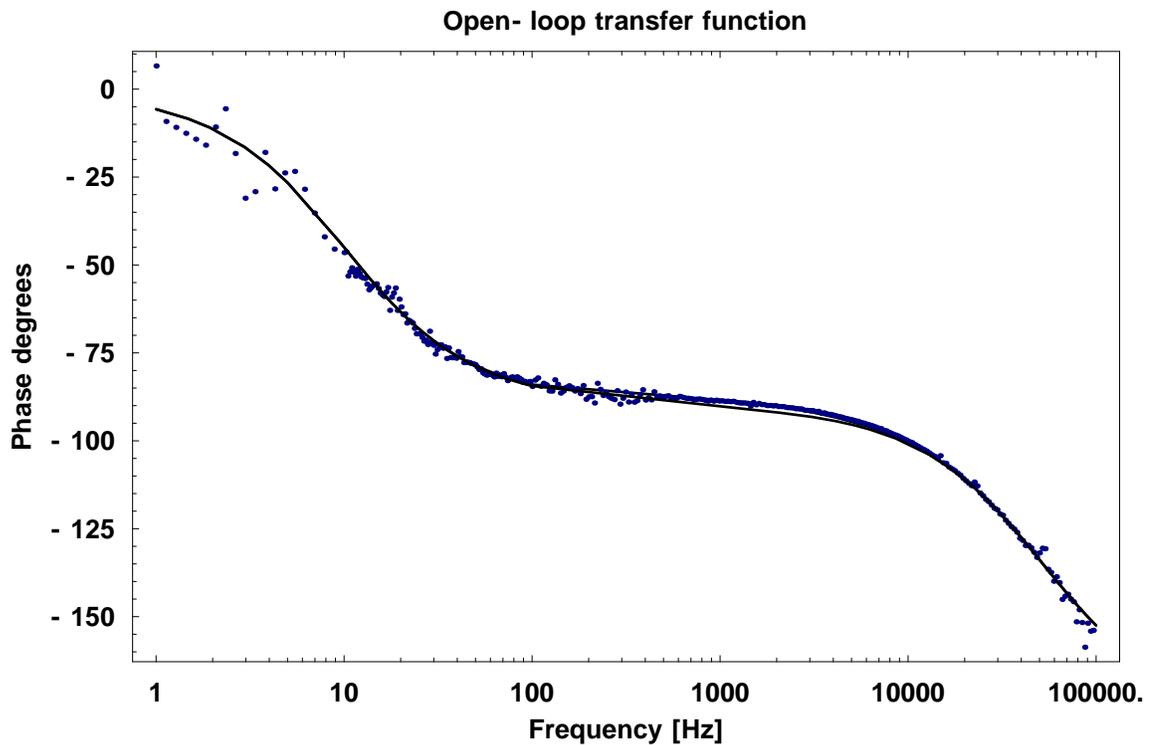

## 3 Results

To measure the photothermal effect, we chop the *Force* beam at various frequencies from 5 Hz to 4 kHz while recording the behavior of the interferometer in the time domain. For each data point, the chopping frequency is set manually. A Tektronix TDS3014 digital oscilloscope records and averages the error signal and photodiode outputs in the time-domain, triggering on the photodiode that monitors the chopper wheel (Fig. 7).

**Figure 7** Scope trace showing 512 averages of measurements of the photothermal effect in the interferometer. In this plot (above the servo's unity gain frequency), the PDH error signal records the mirror's expansion and contraction.

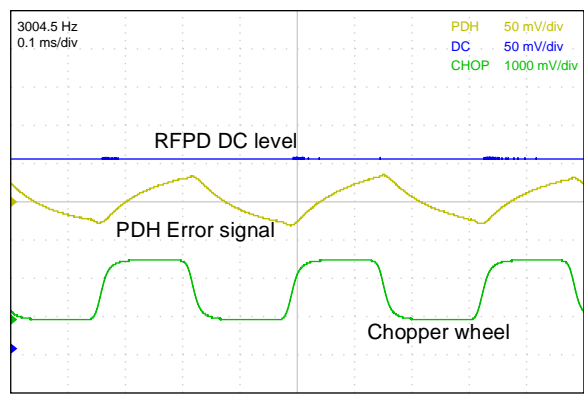

The data are recorded as averages of several short time series, rather than as a single long time series, to reduce the effects of phase noise in the chopper wheel. For most points, 512 samples are averaged. At 20 Hz and below, fewer samples are needed. To set an upper limit on systematic error (See section 4), we also measure the DC output of the photodiode with a digital lock-in amplifier, triggered by the TTL signal from the chopper wheel.

Since the photothermal effect is described by a linear system, we can characterize it by its frequency response. For each measurement frequency, we convolve each channel with sine and cosine functions at the chopper wheel frequency $\omega$, assuming an arbitrary start point, and integrating over an integer number of cycles. From this, we calculate the magnitude response of the photothermal effect in units of Meters/Watt:

$$C_{PDH} \times \frac{\sqrt{\langle \sin(\omega t) \cdot \epsilon(t) \rangle^2 + \langle \cos(\omega t) \cdot \epsilon(t) \rangle^2}}{P_F(\omega)} |H_{R\text{-}D}(f)| \qquad \textbf{(EQ 7)}$$

The numerator in EQ 7 is the magnitude of the Fourier component of the displacement of the mirror at the chopper frequency. Likewise, $P_F(\omega)$ is the Fourier component of the power absorbed by the test mirror at the chopper frequency. $P_F(\omega)$ is the product of the power in the *Force* beam (122 mW), the power absorption of the test mirror (1.33%), the visibility of the IFO (0.5), the finesse of the IFO (37), and a factor of $1/\pi$ (the magnitude of the fundamental mode of a square wave is $1/\pi$ of its peak-to-peak amplitude). The overall scaling of the magnitude data is set by adjusting the absorptivity of the test mirror between 1% and 2%. This method lets us plot the data directly against the models $F_B(\omega)$ and $F_C(\omega)$ by setting $P_{abs} = 1\,W$. This choice of normalization does not affect the phase measurement.

The phase of the photothermal effect is measured by comparing the phase difference between the fundamental frequency components of $\epsilon$ and the chopper signal. The effect of the servo is compensated for by adding the phase of the return-difference function:

$$\text{atan} \frac{\langle \cos(\omega t) \cdot \epsilon(t) \rangle}{\langle \sin(\omega t) \cdot \epsilon(t) \rangle} - \text{atan} \frac{\langle \cos(\omega t) \cdot \text{Chop}(t) \rangle}{\langle \sin(\omega t) \cdot \text{Chop}(t) \rangle} + \text{phase}(H_{R\text{-}D}(f)) \qquad \textbf{(EQ 8)}$$

The predictions for these plots use the properties of Aluminum 6061-T6 shown in Table 1. Poisson's ratio is assumed to be $\sigma = 0.33$, and the thermal conductivity is $\kappa = 140\,W/mK$. Reference values for the thermal conductivity of Al vary considerably, so we analyze a range of values. Fig. 8 shows the magnitude and phase data with the model from Cer-donio *et al.* (EQ 2) and the upper limit set by the Braginsky *et al.* model (EQ 1). The dashed lines are two additional predictions based on EQ 2. Without changing any other parameters, the upper dashed curve in the magnitude plot assumes a thermal conductivity of 130 W/mK, while the lower dashed curve assumes a thermal conductivity of 150 W/mK. Their relative positions are reversed in the phase plot of Fig. 8. Systematic errors in the magnitude data are smaller than the data points. Known systematic uncertainties in the phase data are indicated as vertical lines (where the lines are not obvious, they are the same size as the data points).

**Figure 8** The solid line is the Cerdonio *et al.* model for $\kappa = 140\,\mathrm{W/mK}$. The dashed lines show the same model, with $\kappa = 140 \pm 10\,\mathrm{W/mK}$. The light gray line in the magnitude plot shows the Braginsky *et al.* model, which sets the asymtotic limit at high frequency. The dots show the measured data. Error bars represent phase uncertainty of the chopper wheel's transitions.

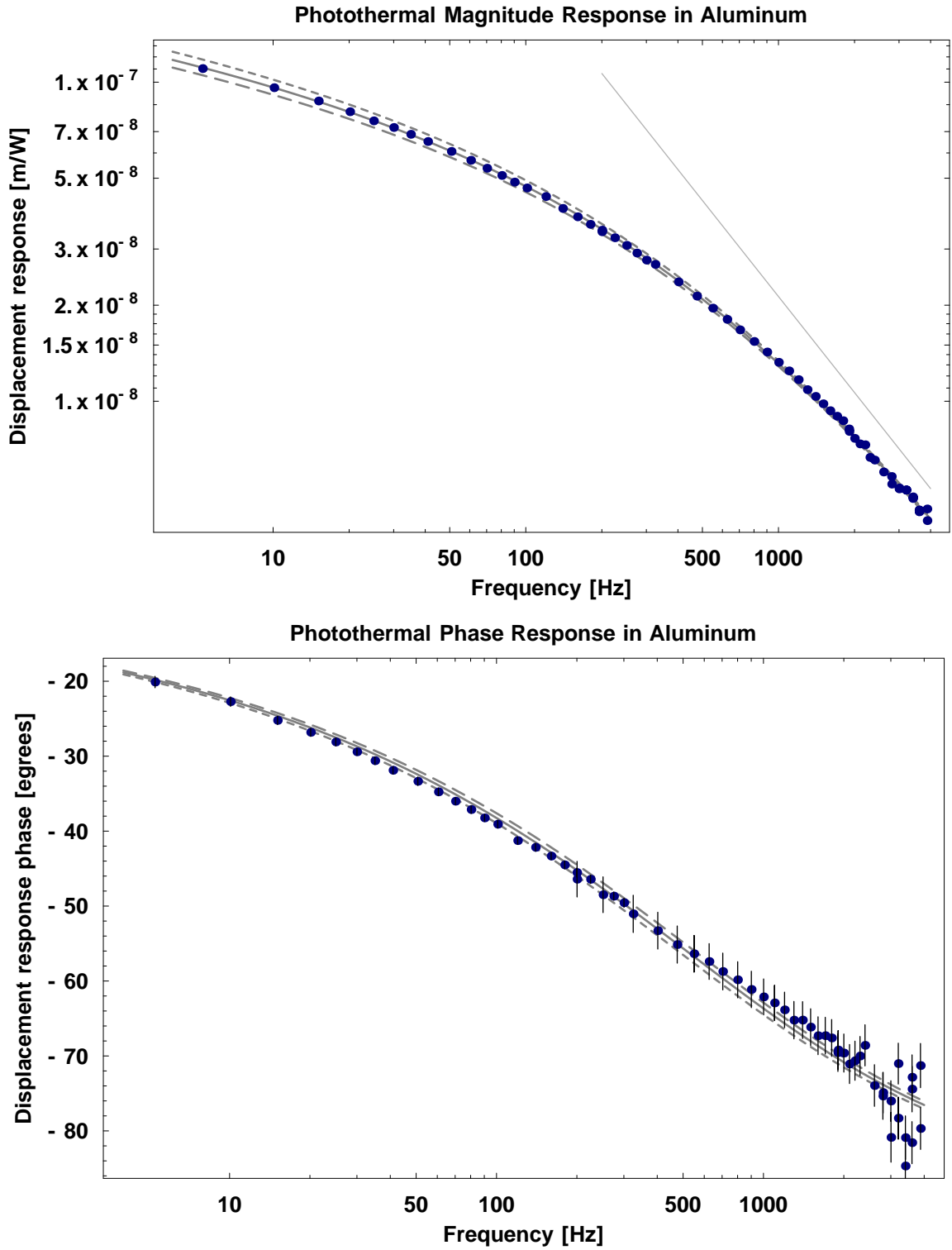

# 4  Error Analysis

- Because of the crosstalk problem, the error signal overestimates the actual length change of the cavity on the order of 1%.

- Above 400 Hz, the gap between the chopper blades is only 10 times the beam diameter. We thus expect to see glitches in the PDH signal as the chopper blade transects the beam, as indicated by the arrow in Fig. 9.

- This also means that, at high frequencies, the phase difference between the PDH and CHOP signals is not well defined.

- The height of the photothermal 'blister' is typically much smaller than the surface roughness of the mirror, so beam distortion effects are negligible.

**Figure 9**  Single data trace. CHOP low means the *Force* beam is blocked. Left: The arrow indicates a jump in the PDH signal. This is the measurement with the most obvious systematic error. Right: a lower frequency measurement shows a non-exponential curve due to the feedback servo.

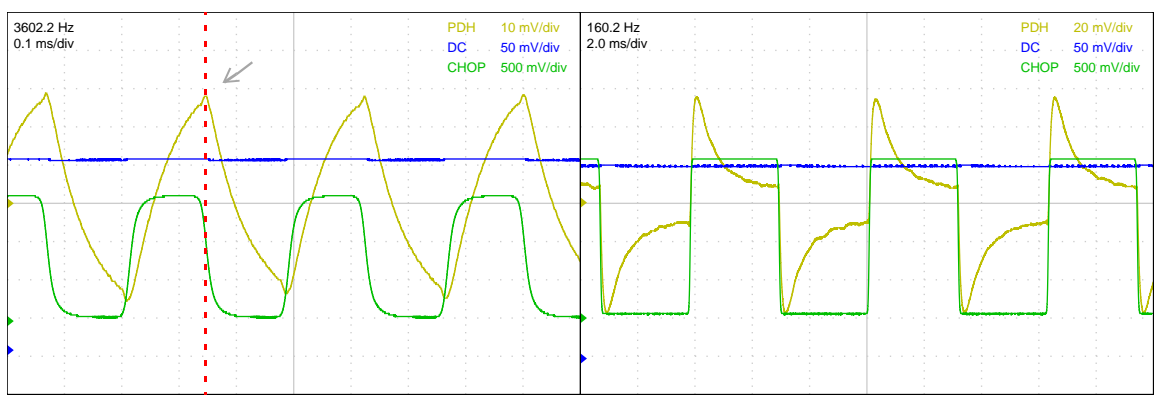

## 4.1  Crosstalk

The light falling on the cavity consists of the *Force* and *Measure* beams. Nominally, these have orthogonal polarizations, and their reflections from the cavity will be separated. In practice, we observe a small amount of each reflecting along the path of the other.

To describe this crosstalk, we define the variable $\zeta^2$ as the power fraction of the light reflected along the wrong path. Since the PDH method measures the light field, not its power, $\zeta$ sets the upper limit on the systematic magnitude error due to interference between the *Force* and *Measure* beams at the RFPD. With the *Force* beam at maximum power (272 mW) and the *Measure* beam blocked after the first PBS, the RFPD observes less than 1 mW of power, which sets the limit $\zeta < 0.06$. We can improve this limit by con-tinuously monitoring the crosstalk as we collect data.

**Figure 10** Left: Expected systematic magnitude error, Right: Measured deviation from model

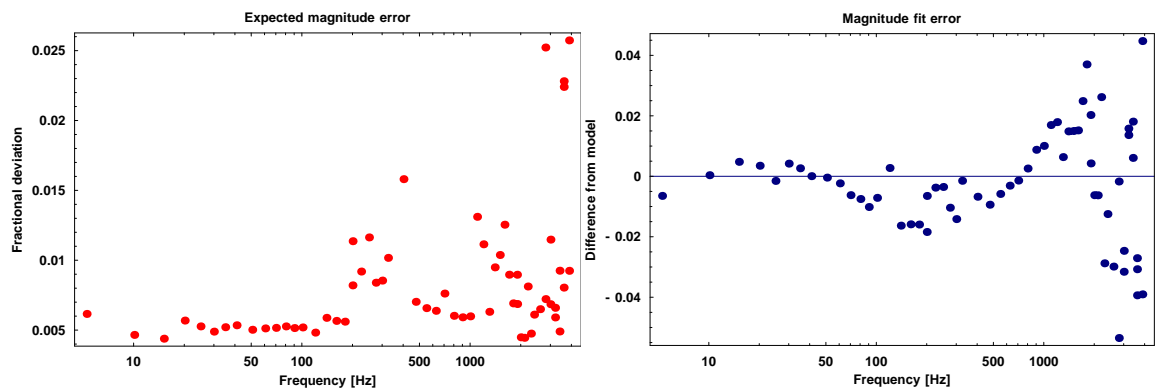

To acquire data, we measure the light reflected along the Measure path with a RF Photo-diode (RFPD) to obtain the error signal. Interference between the *Force* and *Measure* beams causes a small jump in the PDH signal correlated with the chopper wheel, and causes a systematic error in the photothermal measurement. The fractional size of the jump in the length measurement is equivalent to the fractional variation in the DC measurement at the photodiode (see Appendix).

The amplitude variations we measured are less than 1% of the DC value for most measurements, which is smaller than the data points in Fig. 8.

## 4.2  Phase uncertainty

We use different chopper wheels above and below 400 Hz. Below 400 Hz the a chopper wheel has wide gaps, but above 400 Hz, the beam diameter is 0.5mm, while the distance between the blades is 5mm. This leads to the wide CHOP transition in Fig. 11. The two vertical lines in Fig. 11 show the phase relative to the beginning of the time series where the chopper signal passes through its midpoint. The phase difference between these points ought to be 180º, but is actually 175º. We estimate that the systematic error in our calculation of the phase of the photothermal effect is the difference between these angles. Fig. 12 shows the expected systematic phase error for each point and the difference between the phase of the data and the phase of the model. The deviation of the data from the model is within our phase measurement uncertainty.

**Figure 11**

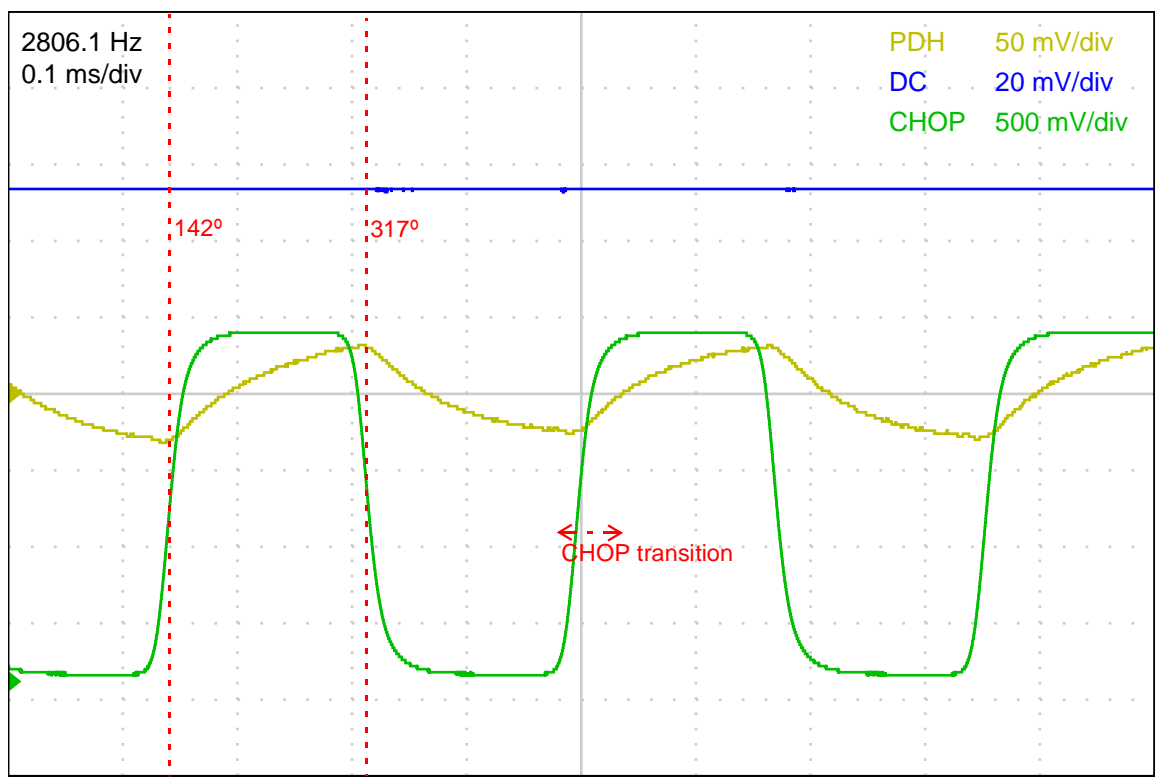

**Figure 12** Left: Expected systematic phase error, Right: Phase difference from measurement

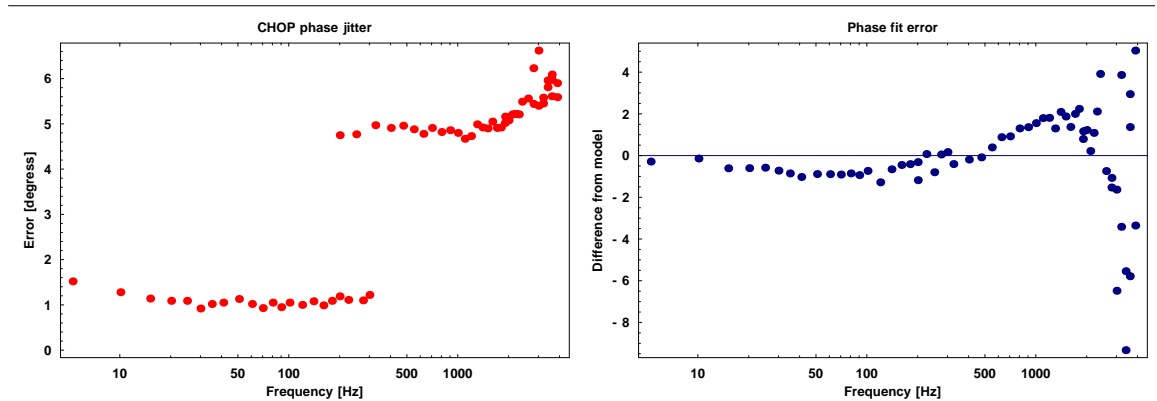

## 4.3 Beam width

Fig. 9 shows the scope trace with the largest chopper glitches in the PDH signal. By inspection, you can see that the area of the bumps indicated by the arrows is less than 1% of the area under the rest of the curve. The glitch is disproportionately represented in the magnitude measurement, which is based on a sine convolution with this curve. The total effect is an overestimate of the magnitude on the order of 1%. Since the glitch is symmetric at high and low peaks, we consider its effect on the phase measurement to be minor.

## 5 Conclusions

Our measurements of the photothermal effect in a solid metal mirror show good agreement with the theories proposed by Braginsky *et al.* and Cerdonio *et al.*, which assume a mirror made from a homogenous, semi-infinite solid. The cross-polarized interferometer appears adequate for measuring optically induced noise in dielectric mirrors.

## Appendix

Consider the case of weak modulation depth at the EOM, where only the carrier and first-order sidebands of the Measure beams are significant. When the laser resonates in the IFO, the light falling on the cavity is

$$P_0 = \left| E_M + E_F \right|^2 \qquad \text{(EQ 9)}$$

$$E_M = \sqrt{P_C}e^{i\omega t} + i\sqrt{P_{SB}}\sin(\Omega t)e^{i\omega t}$$
$$E_F = \sqrt{P_F}e^{i\omega t}e^{i\phi} \qquad \text{(EQ 10)}$$

In general, $\phi$ is not necessarily stationary in time.

**Table 4**

| | |
|---|---|
| $\zeta$ | Fraction of *Force* light with the same polarization as the *Measure* beam |
| $\phi$ | Random phase difference between the *Measure* and *Force* path |
| $P_F, P_{SB}, P_C$ | *Force*, Sideband, Carrier (*Measure* beam) power |
| $\omega$ | Light frequency |
| $\Omega$ | EOM modulation frequency |

Following the method of [Black 2001], and assuming weak sidebands, the *Measure* beam is expanded as

$$E_M = \sqrt{P_C}e^{i\omega t} + \sqrt{P_{SB}/2}(e^{i(\omega+\Omega)t} - e^{i(\omega-\Omega)t}) \qquad \text{(EQ 11)}$$

The amplitude reflection coefficient depends on the input ($r_1$) and 'test' ($r_2$) mirror amplitude reflectivities and input mirror losses ($L_1$). Lower-case letters indicate field coefficients, while capital letters apply to power.

$$F(\omega) = \frac{r_2(1 - L_1)e^{i\omega/FSR} - r_1}{1 - r_1 r_2 e^{i\omega/FSR}}$$

(EQ 12)

The quantity $\omega/FSR$ may be replaced by $4\pi\Delta L/\lambda$ to convert frequency shifts to length changes. In this setup, the geometric mean of $r_1$ and $r_2$ is $\sqrt{0.96}$. That the cavity visibility is 50% on resonance implies that the reflectivities are mismatched or that the input mirror has significant losses.

The field reflected from the IFO is

$$E_{refl} = \sqrt{P_F}e^{i\omega t}e^{i\phi}F(\omega) + \sqrt{P_C}e^{i\omega t}F(\omega) +$$
$$\sqrt{P_{SB}}(e^{i(\omega+\Omega)t}F(\omega+\Omega) - e^{i(\omega-\Omega)t}F(\omega-\Omega))$$

(EQ 13)

The polarizing optics split the *Force* and *Measure* beams on their return paths, and direct the *Measure* beam, and a small amount of the *Force* beam, toward the RFPD. Between the IFO and the photodiode, the light power is attenuated overall by various filters and transmissive optics, so that the field that we measure at the RFPD is proportional to

$$E_{PD} \propto \zeta\sqrt{P_F}e^{i\omega t}e^{i\phi}F(\omega) + \sqrt{P_C}e^{i\omega t}F(\omega) +$$
$$\sqrt{P_{SB}}[e^{i(\omega+\Omega)t}F(\omega+\Omega) - e^{i(\omega-\Omega)t}F(\omega-\Omega)]$$

(EQ 14)

If the demodulation phase at the mixer is chosen to maximize the zero-crossing slope of the PDH signal, then the maximum amount of systematic error due to polarization crosstalk will occur when $\phi \to 0$ or $\pi$. Consider all four fields as phasors. The sum of the carrier and force fields is a field in phase with the carrier. The PDH signal arises from the beat between the carrier (or the sum of the carrier and the noise field) and the two sidebands.

The optical power reaching the photodiode is

$$P_{PD} = |E_{PD}|^2$$

(EQ 15)

After some algebra (again following [Black 2001]), this may be written as

$$P_{PD} \propto \zeta^2 P_F + P_C + 2\zeta\sqrt{P_F P_C}\cos(\phi) + \tag{EQ 16}$$

$$2\sqrt{P_S P_C}\{$$

$$\mathrm{Re}[F(\omega)F^*(\omega+\Omega) - F^*(\omega)F(\omega-\Omega)]\cos\Omega t +$$

$$\mathrm{Im}[F(\omega)F^*(\omega+\Omega) - F^*(\omega)F(\omega-\Omega)]\sin\Omega t\} +$$

$$2\zeta\sqrt{P_F P_S}\{$$

$$\mathrm{Re}[F(\omega)F^*(\omega+\Omega)e^{i\phi} - F^*(\omega)F(\omega-\Omega)e^{-i\phi}]\cos\Omega t +$$

$$\mathrm{Im}[F(\omega)F^*(\omega+\Omega)e^{i\phi} - F^*(\omega)F(\omega-\Omega)e^{-i\phi}]\sin\Omega t\} +$$

$$2\Omega \text{ terms...}$$

The terms proportional to $\sqrt{P_F P_S}$ may be simplified. In the 'linear' region of the PDH signal, near a cavity resonance,

$$F(\omega)F^*(\omega+\Omega) = -F^*(\omega)F(\omega-\Omega) \tag{EQ 17}$$

Below are plots of the magnitude and phase of the terms in EQ 17 for this experiment which demonstrate this. The frequency shifts induced by the photothermal effect in this experiment are less than 400 kHz

**Figure 13** Red: LHS expression, Blue: RHS expression

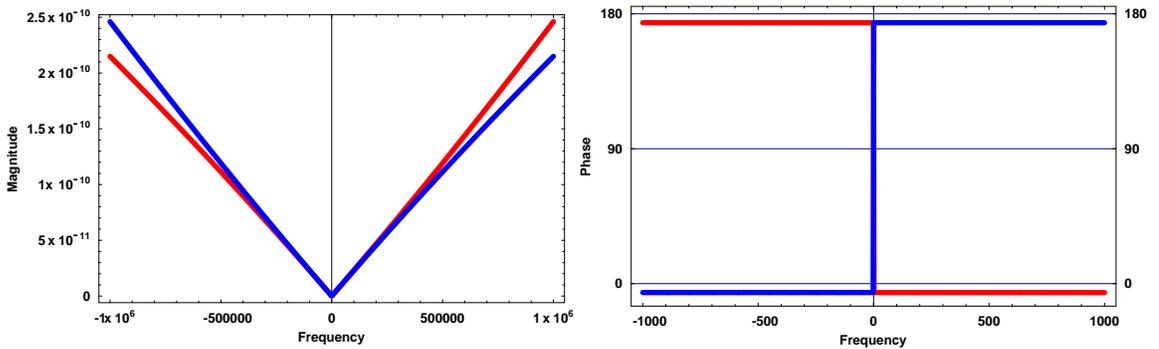

Using this approximation, we can make the replacement $ae^{i\phi} + ae^{-i\phi} = 2a\cos\phi$ and combine terms to make the approximation of EQ 18.

$$P_{PD} \propto (\zeta^2 P_F + P_C + 2\zeta\sqrt{P_F P_C}\cos(\phi))|F(\omega)|^2 + \qquad \text{(EQ 18)}$$

$$2\sqrt{P_S}(\sqrt{P_C} + \zeta\sqrt{P_F}\cos(\phi))\{$$

$$\mathrm{Re}[F(\omega)F^*(\omega + \Omega) - F^*(\omega)F(\omega - \Omega)]\cos\Omega t +$$

$$\mathrm{Im}[F(\omega)F^*(\omega + \Omega) - F^*(\omega)F(\omega - \Omega)]\sin\Omega t\} +$$

$$2\Omega \text{ terms...}$$

We measure high frequency (RF) and low frequency (DC) output of the photodiode separately. Because the mirror reflectivities are not perfectly matched in this experiment, the DC components of EQ 18 are nonzero at resonance.

$$P_{DC} = |F(0)|^2(P_C + 2\zeta\sqrt{P_F P_C}\cos\phi + \alpha^2 P_F) \qquad \text{(EQ 19)}$$

$$F(0) = (r_2 - r_1)/(1 - r_1 r_2) \qquad \text{(EQ 20)}$$

To obtain the error signal from the RF components, we demodulate the RF components $P_{PD}$ at frequency $\Omega$ with a mixer. The low-frequency component of the mixer output, the error signal, is

$$V_{demod} = V_0(\mathrm{Re}[F(\omega)F^*(\omega + \Omega) - F^*(\omega)F(\omega - \Omega)]\cos\theta + \qquad \text{(EQ 21)}$$

$$\mathrm{Im}[F(\omega)F^*(\omega + \Omega) - F^*(\omega)F(\omega - \Omega)]\sin\theta)$$

where $\theta$ is the phase shift between the local oscillator and RF inputs at the mixer and $V_0$ is determined from the OLTF (See page 12). $\theta$ is chosen by trial and error to maximize the slope of the PDH signal on resonance, and measured by a fit to EQ 21 (Fig. 3, Table 3).

The crosstalk factor $\zeta$ is small (around 1%), and $P_F$ and $P_C$ are comparable in magnitude, so the dominant variations in both the error signal and DC signals measured by the photodiode depend on the *Force* beam by the factor $2\zeta\sqrt{P_F/P_C}\cos\phi$.

## References

Bennis 1998                    G. L. Bennis *et al.*, "Thermal diffusivity measurement of solid materials by the pulsed photothermal displacement technique", Journal of Applied Physics 84, 3602-3610 (1998)

Black 2001            Eric D. Black, "An introduction to Pound-Drever-Hall laser frequency stabilization," American Journal of Physics 69, 79-87 (2001).

Braginsky 1999       V. B. Braginsky, M. L. Gorodetsky, S. P. Vyatchanin, "Thermodynamical fluctuations and photo-thermal shot noise in gravitational wave antennae," Physics Letters A 264, 1-10 (1999).

Braginsky 2000       V. B. Braginsky, M. L. Gorodetsky, S. P. Vyatchanin, "Thermo-refractive noise in gravitational wave antennae", Physics Letters A 271, 303-307 (2000)

Cerdonio 2001        M. Cerdonio, L. Conti, A. Heidmann, M. Pinard, "Thermoelastic effects at low temperatures and quantum limits in displacement measurements," Physical Review D 63 082003 (2001). link.aps.org/abstract/PRD/v63/e082003

De Rosa 2002         M. De Rosa, L. Conti, M. Cerdonio, M. Pinard, F. Marin, "Experimental measurement of photothermal effect in Fabry-Perot cavities," www.arXiv.org/gr-qc/0201038 (2002).

Drever 1983          R. W. P. Drever *et al.* "Laser phase and frequency stabilization using an optical resonator," Applied Physics B 31, 91-105 (1983)

Kogelnik 1966        Herwig W. Kogelnik and T. Li, "Laser Beams and Resonators," Applied Optics 5, 1550-1567 (1966).

Liu 2000             Y. T. Liu and K. S. Thorne, "Thermoelastic noise and homogeneous thermal noise in finite sized gravitational-wave test masses," Physical Review D 62, 122002 (2000). http://link.aps.org/abstract/PRD/v62/e122002

Olmstead 1983        M. A. Olmstead, N. M. Amer, S. Kohn, "Photothermal Displacement Spectroscopy: An Optical Probe for Solids and Surfaces," Applied Physics A 32, 141-154 (1983).

1                    Mike Lauer, Research Electro-Optics Inc., personal communication, 1999.

2                    Lightwave Electronics web site:
                     http://www.lwecorp.com/solutions/model_126.htm

CVI Laser web site
http://www.cvilaser.com/Catalog.asp?filename=/optics/bil-
lsroptc-936.asp